\documentclass[runningheads]{llncs}
\usepackage{graphicx}
\usepackage[misc,geometry]{ifsym}
\usepackage{enumitem}
\usepackage{hyperref}
\usepackage{multirow}
\usepackage{array}
\usepackage{subfigure}
\usepackage{booktabs}
\usepackage{xcolor}
\usepackage{amsmath,bm,amssymb}

\begin{document}
\title{Self-Sampling Training and Evaluation for the Accuracy-Bias Tradeoff in Recommendation}
\titlerunning{Self-Sampling Training and Evaluation}
%
\author{Dugang Liu\inst{1,2} \and
Yang Qiao\inst{3} \and
Xing Tang\inst{3} \and
Liang Chen\inst{3} \and
Xiuqiang He\inst{3} \and
Weike Pan\inst{1}\textsuperscript{(\Letter)} \and
Zhong Ming\inst{1}\textsuperscript{(\Letter)}
}
\authorrunning{Liu et al.}
%
\institute{College of Computer Science and Software Engineering, Shenzhen University, Shenzhen, China\\
\email{dugang.ldg@gmail.com,\{panweike,mingz\}@szu.edu.cn} \and
Guangdong Laboratory of Artificial Intelligence and Digital Economy (SZ) \and
FIT, Tencent, Shenzhen, China\\
\email{\{sunnyqiao,shawntang,leocchen,xiuqianghe\}@tencent.com}
}
\maketitle              
\begin{abstract}
Research on debiased recommendation has shown promising results. However, some issues still need to be handled for its application in industrial recommendation. For example, most of the existing methods require some specific data, architectures and training methods. In this paper, we first argue through an online study that arbitrarily removing all the biases in industrial recommendation may not consistently yield a desired performance improvement. For the situation that a randomized dataset is not available, we propose a novel self-sampling training and evaluation (SSTE) framework to achieve the accuracy-bias tradeoff in recommendation, i.e., eliminate the harmful biases and preserve the beneficial ones. Specifically, SSTE uses a self-sampling module to generate some subsets with different degrees of bias from the original training and validation data. A self-training module infers the beneficial biases and learns better tradeoff based on these subsets, and a self-evaluation module aims to use these subsets to construct more plausible references to reflect the optimized model. Finally, we conduct extensive offline experiments on two datasets to verify the effectiveness of our SSTE. Moreover, we deploy our SSTE in homepage recommendation of a famous financial management product called Tencent Licaitong, and find very promising results in an online A/B test.

\keywords{Debiased recommendation \and Self-sampling \and Self-training \and Self-evaluation.}
\end{abstract}
\section{Introduction}\label{sec:intro}
A user will inevitably suffer from various biases during the interaction with a recommender system, which will lead to inherent variability in the feedback data.
As a result, the collected data may not be able to reflect a user's true preferences~\cite{liu2022debiased}.
Ignoring these biases will allow a recommendation model trained based on the feedback data to inherit and even amplify their influence, which is not conducive to the long-term and healthy development of a recommender system.
Therefore, how to reasonably and effectively mitigate the bias problem in the feedback data is an important challenge.

Existing debiased recommendation methods can be mainly divided into two lines, including debiased recommendation with a randomized dataset and without a randomized dataset.
A randomized dataset is collected with a specific uniform policy instead of a recommendation model, which can be regarded as a good unbiased proxy due to the random selection operation used for item assignment~\cite{bonner2018causal}.
With the unbiased information contained in a randomized dataset, the first line focuses on designing different joint training modules to transfer them to a recommendation model trained on a biased feedback data~\cite{bonner2018causal,liu2020general,chen2021autodebias}.
In the case where a randomized dataset is unavailable, the second line mainly ensures the unbiasedness of an optimization objective and guides the design of the model architecture by introducing some theoretical framework~\cite{saito2020unbiased,liu2022debiased,wang2022invariant}.
Although the existing debiased recommendation methods have shown promising results, their application in an industrial recommendation is still lacking sufficient insight since most of them require some specific data, architectures, and training methods.

\begin{figure*}[htbp]
\centering
\includegraphics[width=1.\textwidth]{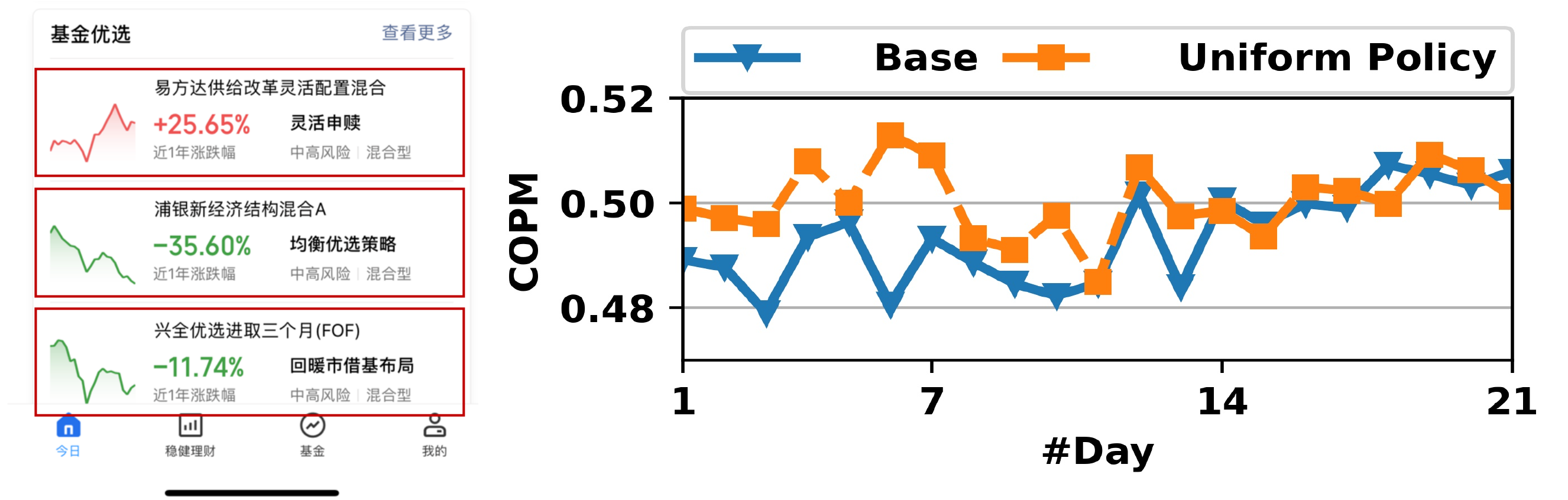}
\caption{A recommendation scenario on the homepage of Tencent Licaitong, and the results of a three-week online study conducted. The contents marked with the red boxes are the recommended items. Note that due to confidentiality, we have transformed the actual COPM values.}
\vspace{-10pt}
\label{fig:1}
\end{figure*}

In particular, an important question that is rarely considered and answered is whether removing all the biases in an industrial recommendation is a desirable goal.
To gain an initial insight into this problem, we first conduct a three-week online study in a real recommendation scenario, where an approximate uniform policy is deployed for comparison with the base model.
This recommendation scenario comes from the homepage recommendation of Tencent Licaitong, which is one of the largest financial recommendation scenarios in China and its display homepage is shown on the left side of Figure~\ref{fig:1}.
Note that more information about this scenario and the evaluation metric COPM can be found in Section~\ref{sec:experiments:online}.
From the right side of Figure~\ref{fig:1}, we can find that the uniform policy will bring an expected performance improvement in the early stage, and this may be due to the unexpected recommendation brought by the random selection operation.
However, in the later stage, the advantage of the uniform policy is not maintained, and degenerates to be similar to the base model.
We argue that this may be due to the fact that the beneficial biases that improve the performance of the base model are removed in the uniform policy, e.g., a high-yield fund product should naturally receive more exposure in this recommendation scenario.
Overall, this means that arbitrarily removing all the biases in an industrial recommendation may not consistently yield a desired performance improvement.

Therefore, in this paper, we propose to use an accuracy-bias tradeoff instead of removing all the biases.
We then propose a simple but effective self-sampling training and evaluation (SSTE) framework to achieve this goal, which preserves the beneficial biases while removing the harmful ones.
Specifically, our SSTE includes three customized modules:
1) a self-sampling module generates some corresponding auxiliary subsets with different degrees of bias from the original training set and validation set;
2) a self-training module combines the original training set and the auxiliary subsets for joint learning to infer the beneficial biases and achieve a better accuracy-bias tradeoff;
3) a self-evaluation module combines the original validation set and the auxiliary subsets to construct a more reasonable reference to better reflect the optimized model offline.
We conduct extensive offline experiments on a public dataset and a real product dataset to verify the effectiveness of our SSTE, including unbiased evaluation and compatibility analysis.
In addition, we also show the strength of our SSTE in an online A/B test.

\section{Related Work}\label{sec:related}
In this section, we briefly review some related works on two research topics, including debiased recommendation and debiased evaluation.

\textbf{Debiased Recommendation.}
According to the types of feedback data involved, existing debiased recommendation methods can be mainly divided into two categories, i.e., debiased recommendation with a randomized dataset and without a randomized dataset.
The former aims to introduce a randomized dataset as an unbiased proxy, and then various ways of its joint training with the original biased feedback data can be designed to exploit the guidance of this unbiased information~\cite{bonner2018causal,liu2020general,chen2021autodebias}.
The latter considers mitigation of the bias problem in the case where a randomized dataset is not available.
The main techniques include assuming and modeling the generation mechanism between a specific bias and some certain features~\cite{marlin2009collaborative,liu2019spiral}, or introducing some theoretical frameworks to construct some corresponding unbiased estimators for this bias problem~\cite{wang2020information,saito2020unbiased,liu2022debiased,wang2022invariant}.
However, most of the existing methods require some specific data, architectures and training methods, which hinders the full exploration and sufficient insights for debiased recommendation in an industrial recommendation.
Our SSTE aims to bridge the gap in this direction.

\textbf{Debiased Evaluation.}
Due to the inherent biases in the feedback data, traditional evaluation metrics may not reflect the real performance of a recommendation model and will lead to a discrepancy between offline and online evaluations.
To solve this problem, most previous works mainly consider from two aspects of measurement design and sample design.
The former aims to design some corresponding unbiased versions for traditional metrics or propose some new unbiased evaluators~\cite{lim2015top,yang2018unbiased,jadidinejad2021simpson}, while the latter focuses on designing some methods that can construct an unbiased validation set~\cite{liang2016causal,bonner2018causal}.
However, most existing methods are inconvenient to application in an industrial recommendation due to the uncontrollable potential risks, i.e., evaluation errors.
Different from them, our SSTE proposes a simple and effective self-evaluation method with the manageable potential risks.
\vspace{-5pt}
\section{The Proposed Framework}\label{sec:framework}
In this paper, we focus on alleviating the bias problem in implicit feedback without a randomized dataset.
Suppose that the training set $\mathcal{D}_{tr}=\{(\bm{x}_i,y_i)\}_{i=1}^{m}$ with $x_i \in \mathcal{X}$ and $y_i \in \mathcal{Y}$ is drawn from a latent distribution $P(\bm{x},y)$, where $m$ is the number of training instances.
$\mathcal{X}=\mathcal{X}_1\times\dots\times\mathcal{X}_d$ is a $d$-dimensional feature space, and $\mathcal{Y} = \left \{0, 1 \right \} $ is a label space.
And the validation set $\mathcal{D}_{val}=\{(\bm{x}_j,y_j)\}_{j=1}^{n}$ is drawn from a latent distribution $Q(\bm{x},y)$, where $n$ is the number of validation instances.
Note that $y=1$ and $y=0$ indicate that a training or validation instance is a positive feedback and a negative feedback, respectively.
\vspace{-5pt}
\subsection{The Accuracy-Bias Tradeoff}\label{sec:framework:tradeoff}
\begin{figure}[htbp]
\centering
    \subfigure[Traditional RS]{
    \label{fig:motivation1}
    \includegraphics[width=0.3\textwidth]{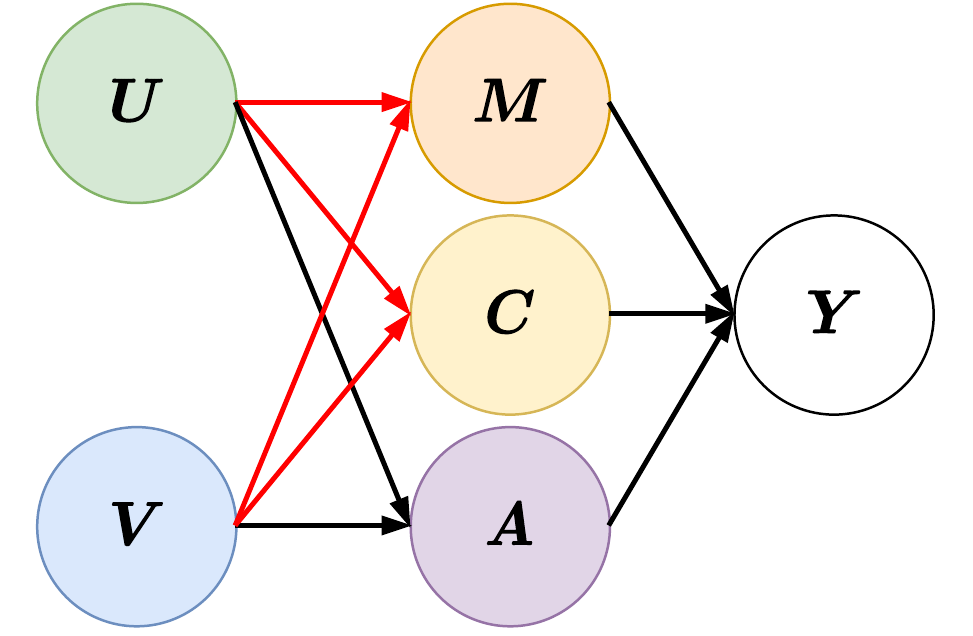}}
    \subfigure[Debiased RS]{
    \label{fig:motivation2}
    \includegraphics[width=0.3\textwidth]{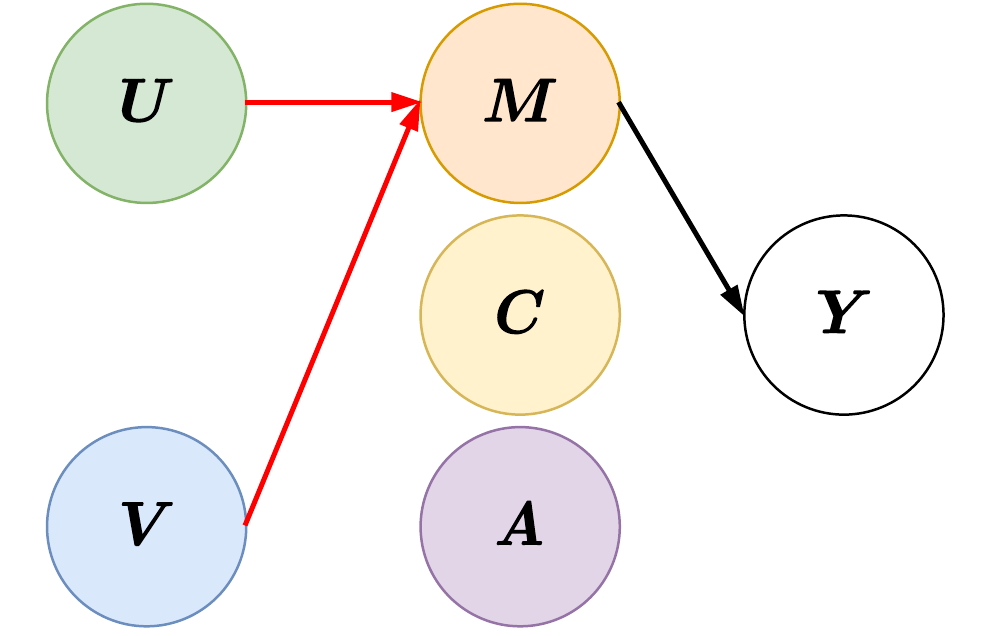}}
    \subfigure[Our Goal]{
    \label{fig:motivation3}
    \includegraphics[width=0.3\textwidth]{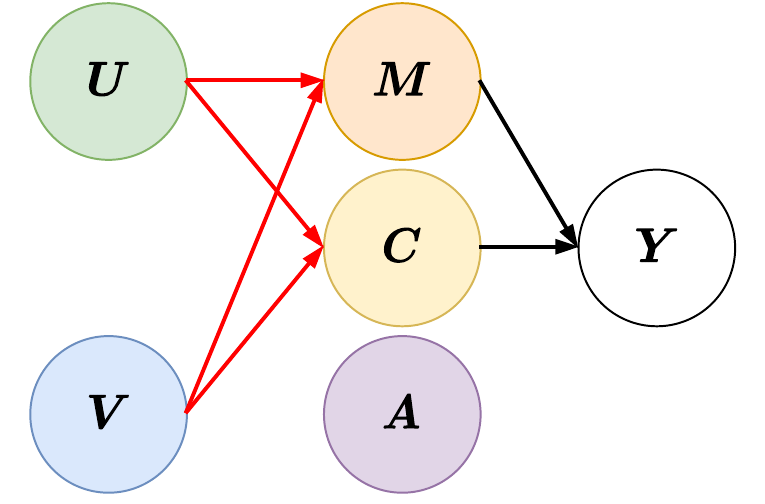}}
\caption{Causal diagrams w.r.t. (a) traditional recommendation, (b) debiased recommendation, and (c) the proposed solution with accuracy-bias tradeoff, where $U$, $V$, $M$, $C$, $A$, and $Y$ denote the users, items, true matching preferences, beneficial bias effects, harmful bias effects, and feedback labels, respectively.}
\vspace{-10pt}
\label{fig:motivation}
\end{figure}

Since the results in Figure~\ref{fig:1} suggest that arbitrarily removing all the biases in an industrial recommendation may not be an ideal choice, we propose a new accuracy-bias tradeoff goal to obtain a more desirable performance improvement, where the key idea is to treat all the biases in the feedback data as a combination of harmful and beneficial ones.
To facilitate the understanding of the difference between our goal and the existing works, we give the causal diagrams of traditional recommendation, debiased recommendation and the proposed new goal in Figure~\ref{fig:motivation}, respectively.
We use $U$, $V$, $M$, $C$, $A$, and $Y$ to denote the users, items, true matching preferences (i.e., $U$'s specific preference for $V$), beneficial bias effects (i.e., the preference offset due to the beneficial biases such as high exposure bias for high-yield funds), harmful bias effects (i.e., the preference offset due to the harmful biases, such as position bias), and feedback labels, respectively.
As shown in Figures~\ref{fig:motivation1} and~\ref{fig:motivation2}, traditional recommendation methods will encode the harmful bias effects, and debiased recommendation methods will remove both the beneficial and harmful bias effects.
From Figure~\ref{fig:motivation3} we can see that unlike them, our goal is to remove the harmful bias effects while retaining the beneficial bias effects.

\vspace{-5pt}
\subsection{Architecture}\label{sec:framework:architecture}
\begin{figure*}[htbp]
    \vspace{-15pt}
    \centering
    \includegraphics[width=1.\linewidth]{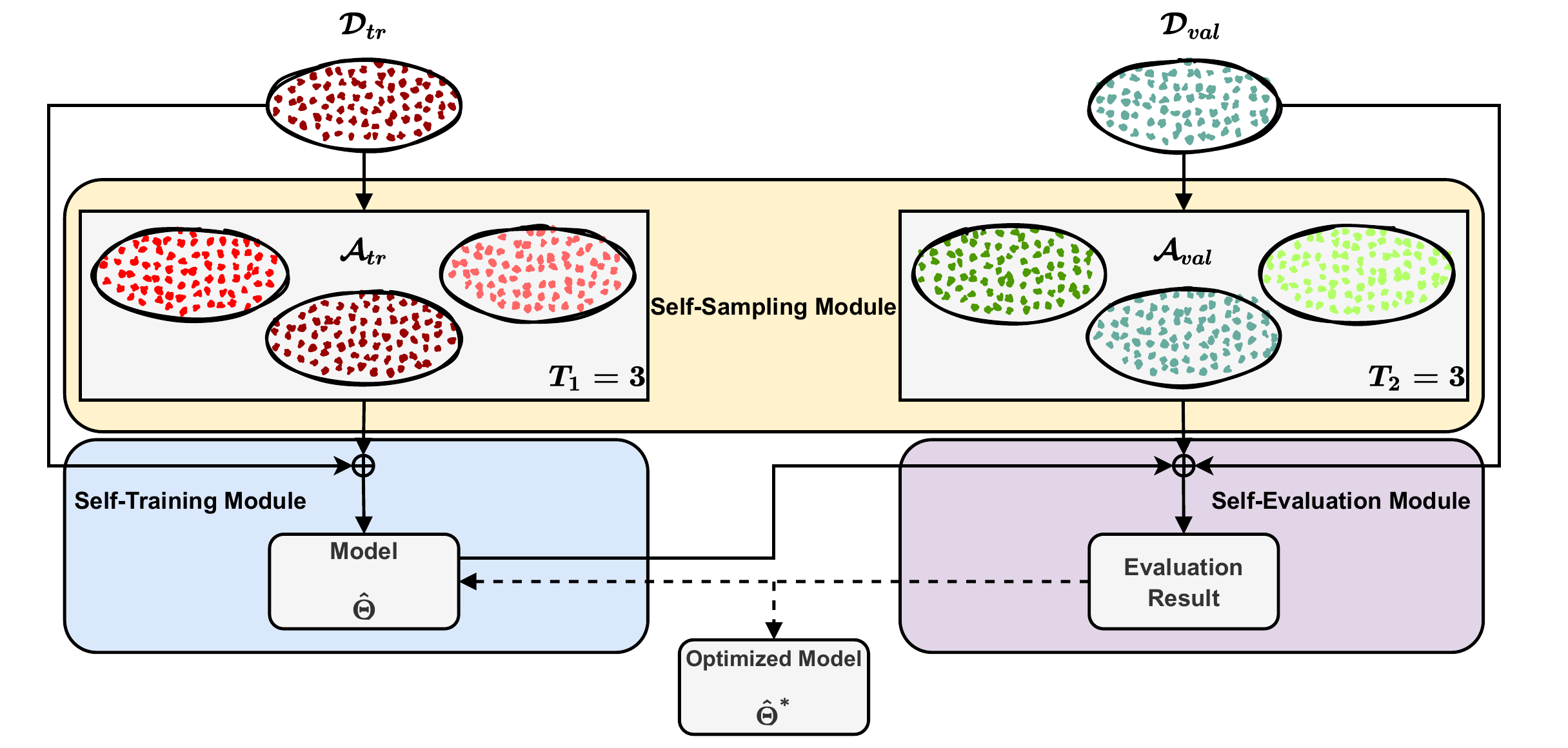}
    \caption{The overall architecture of the proposed SSTE, where the core components are a self-sampling module, a self-training module and a self-evaluation module.}
    \vspace{-30pt}
    \label{fig:framework}
\end{figure*}

We propose a simple but effective self-sampling training and evaluation (SSTE) framework to achieve the desired accuracy-bias tradeoff, where its overall architecture is shown in Figure~\ref{fig:framework}.
Given a training set $\mathcal{D}_{tr}$ and a validation set $\mathcal{D}_{val}$, a self-sampling module constructs a set of auxiliary subsets with different degrees of bias based on $\mathcal{D}_{tr}$ and $\mathcal{D}_{val}$, i.e., $\mathcal{A}_{tr}=\{\hat{\mathcal{D}}^{i}_{tr}\}^{T_1}_{i=1}$ and $\mathcal{A}_{val}=\{\hat{\mathcal{D}}^{i}_{val}\}^{T_2}_{i=1}$.
$\hat{\mathcal{D}}^{i}_{tr}$ is an auxiliary subset sampled from $\mathcal{D}_{tr}$ based on a specific strategy, and $T_1$ is the number of auxiliary subsets equipped for $\mathcal{D}_{tr}$.
$\hat{\mathcal{D}}^{i}_{val}$ and $T_2$ are similarly defined for $\mathcal{D}_{val}$.
Then, a self-training module receives $\mathcal{D}_{tr}$ and $\mathcal{A}_{tr}$, and updates a recommendation model $\hat{\Theta}$ by jointly training with some shared parameters.
The updated model $\hat{\Theta}$ is then passed to the self-evaluation module.
And after combining $\mathcal{D}_{val}$ and $\mathcal{A}_{val}$, the defined new evaluation method is used to obtain the performance corresponding to the current training iteration.
If the convergence condition is not met, the self-training module and the self-evaluation module continue to be executed alternately. 
And once it is met, the optimized recommendation model $\hat{\Theta}^{*}$ will be output.

\subsection{Training}\label{sec:framework:training}
Next, we describe each module in detail based on the training process.

\begin{figure*}[htbp]
    \vspace{-10pt}
    \centering
    \includegraphics[width=0.75\linewidth]{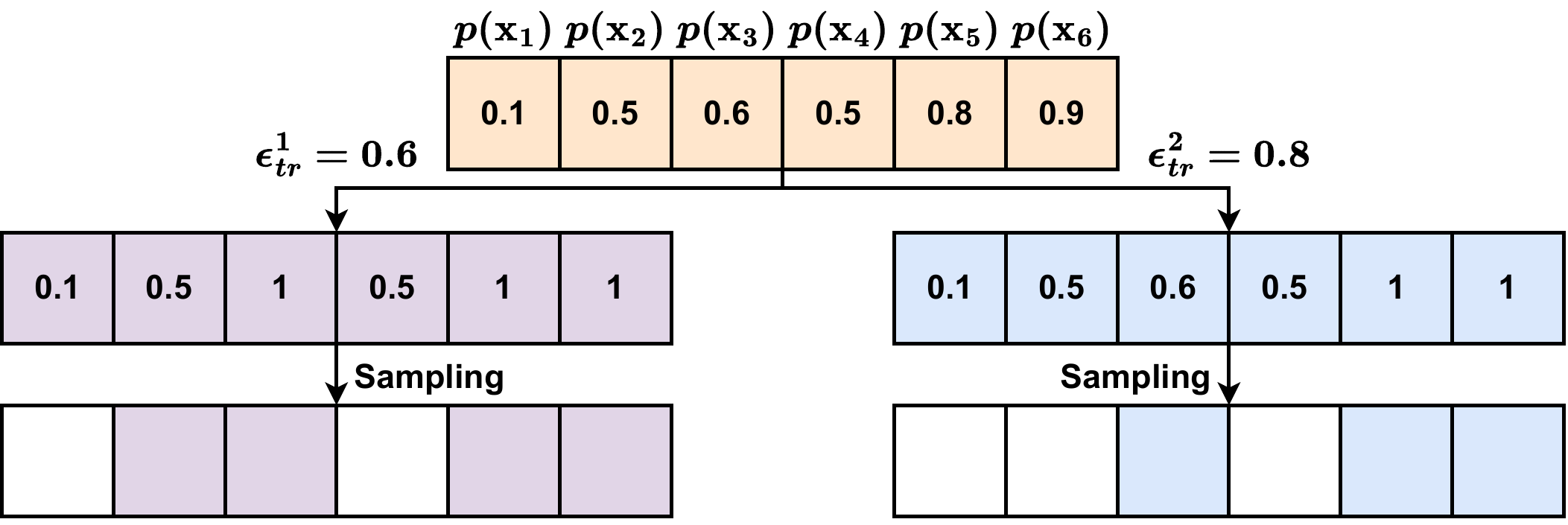}
    \caption{The schematic diagram of a sampling process.}
    \vspace{-10pt}
    \label{fig:sampling}
\end{figure*}

\textbf{The Self-Sampling Module.} 
We propose an auxiliary subset sampling method based on truncated inverse propensity score (tIPS).
Specifically, taking the training instances as an example, we first obtain the sampled probability of each instance $p(\bm{x}_i)$ by some existing IPS estimators in debiased recommendation, where a higher $p(\bm{x}_i)$ means this sample is more important for debiasing.
We then choose a set of different truncation thresholds $\{\epsilon^i_{tr}\}^{T_1}_{i=1}$ and use each threshold separately to adjust the obtained $p(\bm{x}_i)$, i.e., keep $p(\bm{x}_i)$ when $p(\bm{x}_i)<\epsilon^i_{tr}$, otherwise modify $p(\bm{x}_i)$ to 1.
This means that we control the level of debiasing by applying more protection to different proportions of important samples.
Finally, based on the modified sampling probabilities, we can obtain a set of auxiliary subsets of $\mathcal{D}_{tr}$, i.e., $\mathcal{A}_{tr}=\{\hat{\mathcal{D}}^{i}_{tr}\}^{T_1}_{i=1}$, where each auxiliary subset corresponds to a truncation threshold.
Similarly, by setting $\{\epsilon^i_{val}\}^{T_2}_{i=1}$ for the validation instances, we can obtain $\mathcal{A}_{val}=\{\hat{\mathcal{D}}^{i}_{val}\}^{T_2}_{i=1}$ corresponding to $\mathcal{D}_{val}$.
The idea behind this sampling operation is to simulate the auxiliary subsets with different degrees of bias and different sets of biases based on the feedback data itself.
Note that the self-sampling module can be executed only once as a preprocessing operation, or it can be re-executed after each round of training.
We give an example of the sampling process in Figure~\ref{fig:sampling}.

\textbf{The Self-Training Module.}
In order to combine $\mathcal{D}_{tr}$ and $\mathcal{A}_{tr}$ to infer the beneficial bias information and constrain the model to achieve a better accuracy-bias tradeoff, we can use any model architecture with some shared parameters for training.
Formally, let $\widetilde{\Theta}=\{\widetilde{\theta},\theta_s\}$ be the model parameters related to $\mathcal{D}_{tr}$, and $\hat{\Theta}=\{\hat{\theta},\theta_s\}$ be the model parameters related to $\mathcal{A}_{tr}$, where $\theta_s$ is the shared parameter.
The final optimization objective function of our SSTE can be expressed as follows,
\begin{equation}\label{equ:1}
\mathop{\min}\limits_{\widetilde{\Theta},\hat{\Theta}} \mathcal{L}_{SSTE} = \mathcal{L}_{\mathcal{D}_{tr}}\left(f\left(\widetilde{\Theta},\bm{x}_i\right),y_i\right) + \mathcal{L}_{\mathcal{A}_{tr}}\left(g\left(\hat{\Theta},\bm{x}_j\right),y_j\right) + \lambda \lVert\widetilde{\Theta}\rVert + \lambda \lVert\hat{\Theta}\rVert,
\end{equation}
where $(\bm{x}_i,y_i)\in\mathcal{D}_{tr}$, $(\bm{x}_j,y_j)\in\mathcal{A}_{tr}$, $f(\cdot)$ and $g(\cdot)$ are the mapping functions, and $\lambda$ and $\lVert\cdot\rVert$ are the tradeoff weight and the regularization terms, respectively.
Note that we will adopt the model $\hat{\Theta}$ during the inference phase.
An intuitive interpretation for Eq.~\eqref{equ:1} is that joint training on data that simulates different environments forces the model to implicitly distinguish all the biases, where those biases that reach more consensus and encode more information are more likely to be beneficial, since the performance gains they bring are more robust across different environments.

\textbf{The Self-Evaluation Module.}
To better capture the optimal model $\hat{\Theta}^{*}$ offline, we propose a bias-robust evaluation method.
Specifically, let the performance of the model $\hat{\Theta}$ on $\mathcal{D}_{val}$ and $\hat{\mathcal{D}}^{i}_{val}$ be $e(\mathcal{D}_{val})$ and $e(\hat{\mathcal{D}}^{i}_{val})$, respectively, at the $i$-th training iteration, where $e(\cdot)$ denotes the main metric adopted.
Combining $\mathcal{D}_{val}$ and $\mathcal{A}_{val}=\{\hat{\mathcal{D}}^{i}_{val}\}^{T_2}_{i=1}$, we can compute the maximum difference $\alpha$ in performance between any two sets,
\begin{equation}\label{equ:2}
\alpha=\mathop{\max}\{\mathop{\max}|e(\mathcal{D}_{val})-e(\hat{\mathcal{D}}^i_{val})|,\mathop{\max}|e(\hat{\mathcal{D}}^i_{val})-e(\hat{\mathcal{D}}^j_{val}|)\},
\end{equation}
where $|\cdot|$ denotes an absolute value operation. 
Since $\mathcal{D}_{val}$ and $\mathcal{A}_{val}$ are simulated for different bias environments, and an ideal optimization model should have a stable performance in different environments, an intuitive idea is that the model with a smaller value of $\alpha$ is better.
Finally, depending on whether $e(\cdot)$ is a higher-better metric, we use $e(\mathcal{D}_{val})-\alpha$ or $e(\mathcal{D}_{val})+\alpha$ as a modified performance result to better capture the optimal model $\hat{\Theta}^{*}$ offline with a manageable risk.
\vspace{-10pt}
\subsection{Remarks}\label{sec:framework:remarks}
IPS-based methods are an important branch in debiased recommendation, but the performance of these methods heavily depends on the estimation accuracy of IPS.
Different from them, our SSTE only uses IPS as a reference to simulate different bias environments, and thus has a greater tolerance for the estimation accuracy of IPS.
Our SSTE can be applied to many industrial recommendation scenarios because it does not depend on a specific data, architecture or training method, and is compatible with most industrial recommendation models.
Furthermore, since the sampled auxiliary subset usually has a much smaller size than the original data, i.e., $|\hat{\mathcal{D}}^{i}_{tr}|\ll m$ and $|\hat{\mathcal{D}}^{i}_{val}|\ll n$, our SSTE does not increase the time and resource overhead too much.

\vspace{-10pt}
\section{Experiments}\label{sec:experiments}
In this section, we first introduce the experimental setup, and then conduct extensive empirical studies and show the effectiveness of our SSTE.
\vspace{-10pt}
\subsection{Experimental Setup}\label{sec:experiments:setup}
\textbf{Datasets.} 
We use a very common benchmark dataset and a real product dataset in our experiments, i.e., Yahoo! R3~\cite{marlin2009collaborative} and Product.
Following the settings of most previous works~\cite{liu2021mitigating,chen2021autodebias}, for Yahoo! R3, we first binarize each rating, where those greater than 3 are denoted as $y=1$, and the rest as $y=0$.
For the biased feedback subset in Yahoo! R3, we randomly divide each user's feedback into training and validation sets with a ratio of $8:2$.
The randomized feedback subset in Yahoo! R3 is all used for unbiased evaluation, since it can be considered as the feedback generated by an unbiased scene.
Product is a subset sampled from the log data collected in Tencent Licaitong's homepage recommendation business, involving about 2.8 million users, 560 items, and 9.6 million feedback.
According to the different properties of the feedback data, we divide them into a biased feedback subset and a randomized feedback subset.
After chronological ordering of the biased feedback subset, we obtain the training and validation sets in a ratio of $8:2$.
Similarly, the whole randomized feedback subset is used for unbiased evaluation.
The statistics of the datasets are shown in Table~\ref{tab:datasets}.
\begin{table}[htbp]
\caption{Statistics of the datasets. P/N represents the ratio between the numbers of positive and negative feedback.}
\centering
\scalebox{0.85}{
\begin{tabular}{c|cc|cc}
\specialrule{0.1em}{3pt}{3pt}
 & \multicolumn{2}{c|}{\textbf{Yahoo! R3}} & \multicolumn{2}{c}{\textbf{Product}} \\
 & \#Feedback & P/N (\%)& \#Feedback & P/N (\%)\\
\specialrule{0.05em}{3pt}{3pt}
training set &254,713 &67.02 &7,133,519 &5.21\\
validation set &56,991 &67.00 &1,633,716 &5.20\\
test set &54,000 &9.64 &870,158 &4.34\\
\specialrule{0.1em}{3pt}{3pt}
\end{tabular}}
\vspace{-15pt}
\label{tab:datasets}
\end{table}

\textbf{Evaluation Metrics.} 
For all the experiments, we employ four evaluation metrics that are widely used in recommender systems, including the area under the ROC curve (AUC), precision (P@K), recall (R@K) and normalized discounted cumulative gain (nDCG).
We report the results of P@K and R@K when K is 5 and 10, and the results of nDCG when K is 50.
Since AUC is one of the most common metrics in an industrial recommendation, we choose it as our main evaluation metric, which is used to search for the best hyperparameters for all the candidate methods.

\textbf{Baselines.} 
To conduct a comprehensive evaluation of our SSTE, in the experiments, we select a set of representative debiased recommendation methods based on only a biased data (i.e., without a randomized dataset), including IPS~\cite{schnabel2016recommendations}, SNIPS~\cite{swaminathan2015self}, CVIB~\cite{wang2020information}, AT~\cite{saito2020asymmetric}, Rel~\cite{saito2020unbiased} and DIB~\cite{liu2022debiased}.
Furthermore, similar to most of the previous works, we adopt matrix factorization (MF)~\cite{koren2009matrix} and neural collaborative filtering (NCF)~\cite{he2017neural} as two backbone models.
Therefore, each of these baselines has two corresponding versions based on different backbones.
Note that we do not include CausE~\cite{bonner2018causal}, KDCRec~\cite{liu2020general} and AutoDebias~\cite{chen2021autodebias} because they require a randomized dataset for bias reduction.

\textbf{Implementation Details.}
All the candidate methods have been implemented on TensorFlow.
We set the optimizer to Adam, and use the hyperparameter search library \textit{Optuna}~\cite{akiba2019optuna} to speed up the hyperparameter search process with AUC as the target on the validation set.
For our SSTE, we set both the number of self-samples $T_1$ and $T_2$ to 1, considering the tradeoff of performance and resource overhead.
To avoid overfitting to the training set, we also employ an early stopping setting with a patience of 5.
We tune the embedding dimension, the regularization weight, the batch size and the learning rate in the range of $\left\{5,10,\cdots,195,200\right\}$, $\left\{1e^{-5},1e^{-4},\cdots,1e^{-1},1\right\}$, $\left\{2^{7},2^{8},\cdots,2^{13},2^{14}\right\}$ and $\left\{1e^{-4},5e^{-4}\cdots5e^{-2},1e^{-1}\right\}$, respectively.
Note that the source codes are available at \url{https://github.com/dgliu/DASFAA23_SSTE}.

\subsection{Overall Results}\label{sec:experiments:overall}
The comparison results between our SSTE and the baselines are shown in Table~\ref{tab:comparison_result}.
When using matrix factorization as the backbone model, as shown in the upper part of Table~\ref{tab:comparison_result}, our SSTE consistently outperforms all the baselines on all the metrics across the two datasets of Yahoo! R3 and Product.
We can also observe that most debiasing methods can improve the unbiased performance of recommendation models to some extent, among which Rel and DIB are the two most competitive baselines.
Compared with them, our SSTE can benefit from self-training and self-evaluation modules to achieve a better result.
When using neural collaborative filtering as the backbone model, as shown in the lower part of Table~\ref{tab:comparison_result}, our SSTE outperforms all the baselines in most cases.
We can find that although our SSTE is slightly weaker than DIB on P@10 and R@10 on Product, it still has a clear advantage on other metrics, especially the main metric AUC.
This may be due to the impact caused by only considering AUC in parameter tuning.
Overall, our SSTE has a better unbiased performance.

\begin{table}[htbp]
\vspace{-10pt}
\caption{Comparison results of unbiased evaluation, where the best results and the second best results are marked in bold and underlined, respectively. AUC is the main evaluation metric.}
\centering
\scalebox{0.85}{
\begin{tabular}{c|cccccc|cccccc}
\specialrule{0.1em}{3pt}{3pt}
 & \multicolumn{6}{c|}{\textbf{Yahoo! R3}} & \multicolumn{6}{c}{\textbf{Product}} \\
\specialrule{0.05em}{3pt}{3pt}
\textbf{Method}& \textbf{AUC} & \textbf{nDCG} & \textbf{P@5} & \textbf{P@10} & \textbf{R@5} & \textbf{R@10}& \textbf{AUC} & \textbf{nDCG} & \textbf{P@5} & \textbf{P@10} & \textbf{R@5} & \textbf{R@10}\\ 
\specialrule{0.05em}{3pt}{3pt}
MF & {.7101} & {.0447} & {.0080} & {.0074} & {.0236} & {.0446} & {.8477} & {.0696} & {.0145} & {.0180} & {0687} & {.1707}\\
\specialrule{0.05em}{3pt}{3pt}
IPS-MF & {.7128} & {.0313} & {.0033} & {.0035} & {.0088} & {.0202} & {.8507} & {.0684} & {0139} & {.0180} & {.0663} & {.1717}\\
SNIPS-MF & {.7101} & {.0334} & {.0049} & {.0048} & {.0131} & {.0271} & {.8509} & {.0684} & {.0140} & {.0181} & {.0670} & {.1717}\\
\specialrule{0.05em}{3pt}{3pt}
CVIB-MF & {.7048} & {.0479} & {.0089} & {.0069} & {.0267} & {.0413} & {.8279} & {.0705} & {.0146} & {.0185} & {.0695} & {.1761}\\
AT-MF & {.7314} & {.0663} & {.0108} & {.0100} & {.0373} & {.0676} & {.8389} & {.0649} & {.0137} & {.0169} & {.0649} & {1595}\\
Rel-MF & {.7440} & {.0835} & {.0151} & {.0131} & {.0508} & {.0859} & \underline{.8519} & \underline{.0785} & \underline{.0177} & {.0199} & \underline{.0838} & {.1882}\\
DIB-MF & \underline{.7547} & \underline{.0920} & \underline{.0169} & \underline{.0145} & \underline{.0538} & \underline{.0930} & {.8510} & {.0781} & {.0159} & \underline{.0202} & {.0764} & \underline{.1922}\\
\specialrule{0.05em}{3pt}{3pt}
SSTE-MF & \textbf{.7591} & \textbf{.0981} & \textbf{.0181} & \textbf{.0153} & \textbf{.0612} & \textbf{.0999} & \textbf{.8525} & \textbf{0878} & \textbf{.0222} & \textbf{.0211} & \textbf{.1062} & \textbf{.2010}\\
\specialrule{0.1em}{3pt}{3pt}


\specialrule{0.1em}{3pt}{3pt}
 & \multicolumn{6}{c|}{\textbf{Yahoo! R3}} & \multicolumn{6}{c}{\textbf{Product}} \\
\specialrule{0.05em}{3pt}{3pt}
\textbf{Method}& \textbf{AUC} & \textbf{nDCG} & \textbf{P@5} & \textbf{P@10} & \textbf{R@5} & \textbf{R@10}& \textbf{AUC} & \textbf{nDCG} & \textbf{P@5} & \textbf{P@10} & \textbf{R@5} & \textbf{R@10}\\ 
\specialrule{0.05em}{3pt}{3pt}
NCF & {.7244} & {.0294} & {.0022} & {.0026} & {.0068} & {.0147} & {.7930} & {.0795} & {.0192} & {.0195} & {.0907} & {.1845}\\
\specialrule{0.05em}{3pt}{3pt}
IPS-NCF & {.7229} & {.0291} & {.0031} & {.0036} & {.0084} & {.0199} & {.8024} & {.0802} & {.0187} & {.0188} & {.0886} & {.1781}\\
SNIPS-NCF & {.7224} & {.0314} & {.0034} & {.0034} & {.0091} & {.0183} & {.8026} & {.0869} & {.0195} & {.0206} & {.0921} & {.1945}\\
\specialrule{0.05em}{3pt}{3pt}
CVIB-NCF & {.7250} & {.0393} & {.0053} & {.0044} & {.0150} & {.0250} & {.8034} & {.0776} & {.0170} & {.0193} & {.0805} & {.1831}\\
AT-NCF & {.7167} & {.0351} & {.0056} & {.0048} & {.0157} & {.0271} & {.7967} & {.0741} & {.0181} & {.0179} & {.0856} & {.1688}\\
Rel-NCF & {.6895} & {.0505} & {.0083} & {.0073} & {.0246} & {.0451} & {.7970} & {.0827} & {.0183} & {.0198} & {.0869} & {.1880}\\
DIB-NCF & \underline{.7454} & \underline{.0671} & \underline{.0114} & \underline{.0096} & \underline{.0365} & \underline{.0602} & \underline{.8049} & \underline{.1011} & \underline{.0233} & \textbf{.0241} & \underline{.1105} & \textbf{.2293}\\
\specialrule{0.05em}{3pt}{3pt}
SSTE-NCF & \textbf{.7561} & \textbf{.0696} & \textbf{.0115} & \textbf{.0101} & \textbf{.0370} & \textbf{.0638} & \textbf{.8061} & \textbf{.1033} & \textbf{.0247} & \underline{.0228} & \textbf{.1175} & \underline{.2171}\\
\specialrule{0.1em}{3pt}{3pt}
\end{tabular}}
\vspace{-25pt}
\label{tab:comparison_result}
\end{table}

\subsection{Compatibility Analysis}\label{sec:experiments:compatibility}
\begin{figure}[htbp]
\vspace{-10pt}
\centering
    \subfigure[MF]{
    \label{fig:compatibility_MF}
    \includegraphics[width=0.487\textwidth]{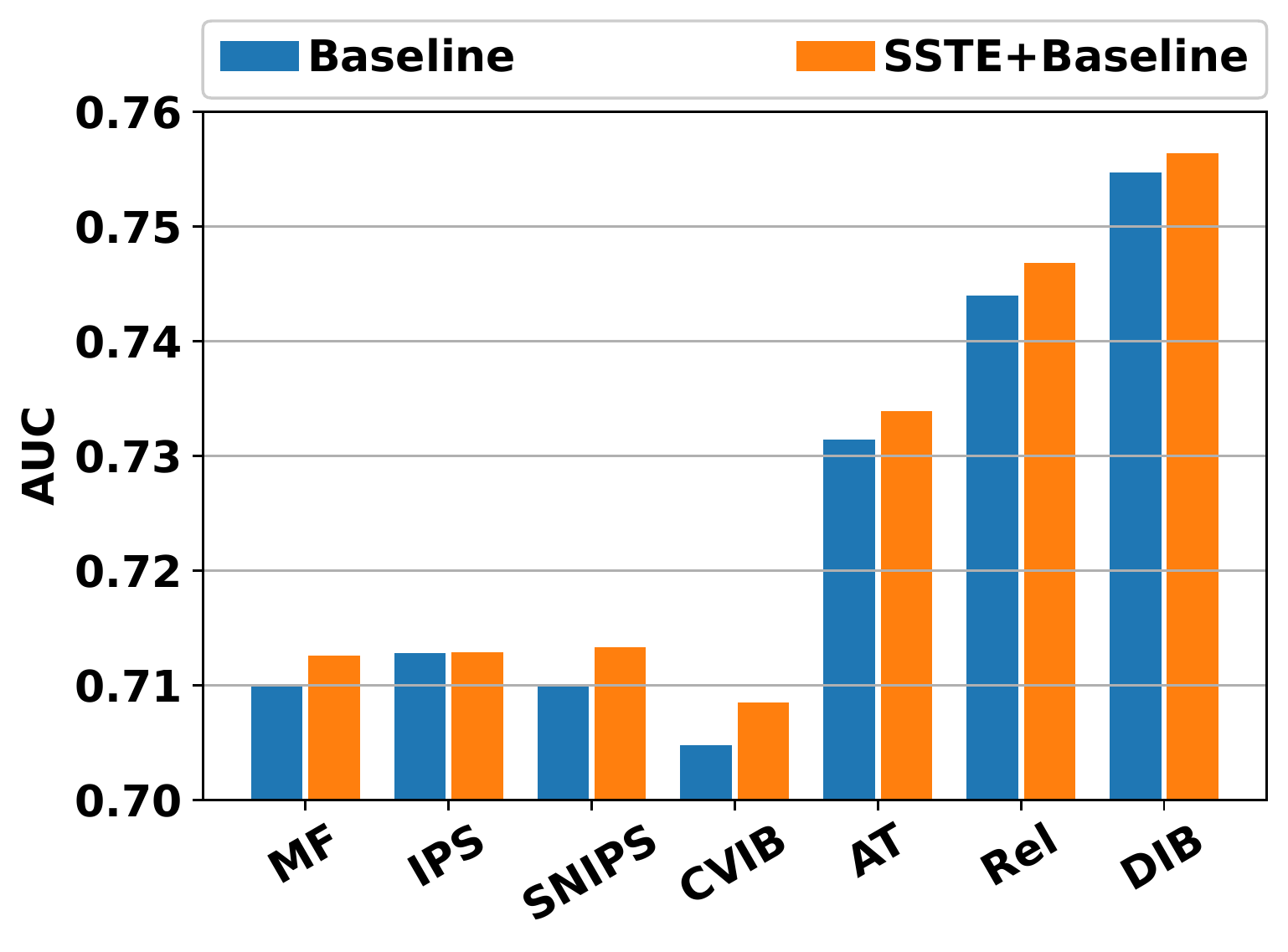}}
    \subfigure[NCF]{
    \label{fig:compatibility_NCF}
    \includegraphics[width=0.487\textwidth]{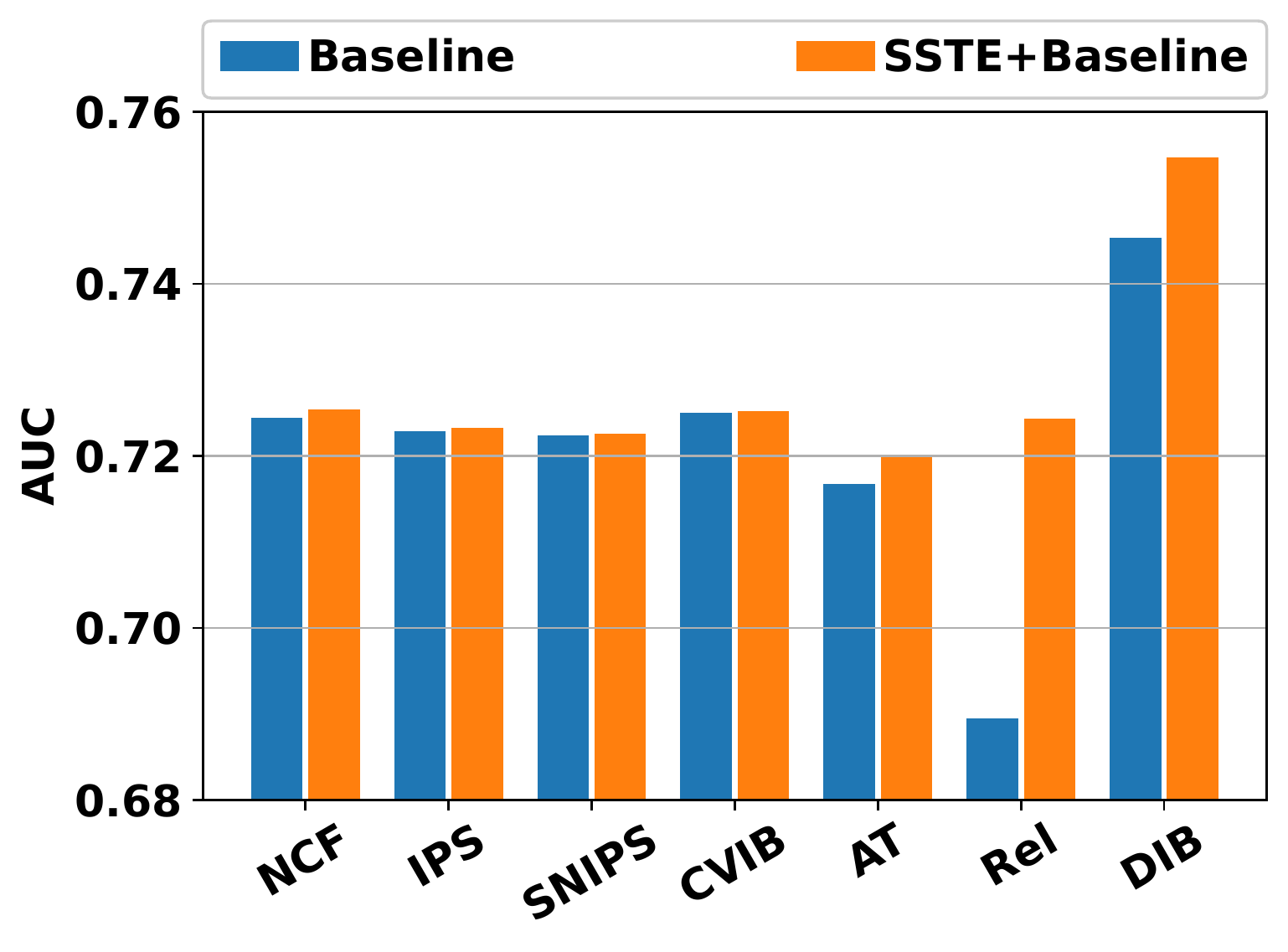}}
\caption{Recommendation results of our SSTE with different baselines on Yahoo! R3.}
\vspace{-30pt}
\label{fig:compatibility}
\end{figure}

As described in Section~\ref{sec:framework}, since our SSTE does not depend on a specific architecture or training method, it can be easily integrated with existing debiased recommendation methods and traditional recommendation methods.
To verify the compatibility of our SSTE, in our experiments, we integrate it with all the baselines and compare it with the original baselines after re-searching for the best hyperparameters.
We report the results on Yahoo! R3 in Figure~\ref{fig:compatibility}.
We can find that our SSTE can bring a significant improvement over the unbiased performance of different baselines in most cases.
In particular, we can observe that the positive effect of our SSTE will be weakened on the debiased recommendation method based on inverse propensity score (IPS).
This may be due to the fact that the inaccurate estimation of IPS would bring an irreconcilable hurt to a recommendation model.

\subsection{Online A/B Test}\label{sec:experiments:online}
\begin{figure*}[htbp]
\vspace{-10pt}
\centering
\includegraphics[width=0.9\textwidth]{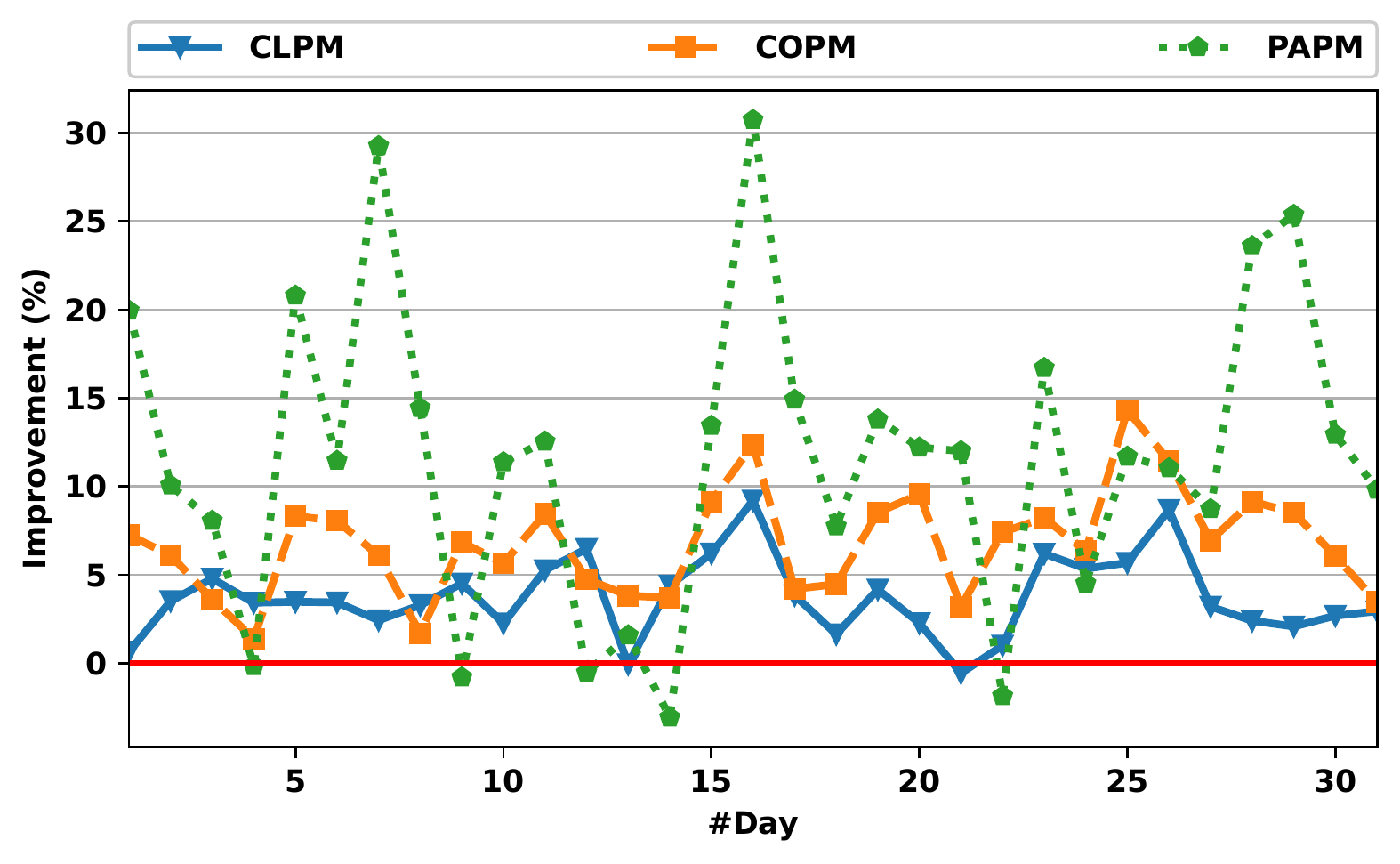}
\caption{The improvement of our SSTE compared with the base model in the online A/B test, including total clicks per mille (CLPM), total conversions per mille (COPM) and purchase amount per mille (PAPM).}
\vspace{-35pt}
\label{fig:online}
\end{figure*}

Finally, we deploy our SSTE in Tencent Licaitong, which is a large-scale internet financial platform dedicated to providing high-quality fund sales services for users.
In this platform, there are tens of millions of active users every day, and a large number of feedback logs are generated and recorded. We conduct online A/B test for one month in the homepage recommendation scenario, which is the first page after user log in.  
In this recommendation scenario, a set of funds will be recommended to the user, and the user can perform some related operations, such as skip, click and purchase. The display page is shown in the left side of Figure~\ref{fig:1}.
The base model compared in the online test is a carefully tuned deep multi-task model in which clicks, conversions, and purchase amounts are predicted separately.
We deploy SSTE on the same architecture, and both models are trained over the same training dataset, which contain more than 300 million logged feedback spanning two months.
For online serving, 10\% users are randomly selected as the experimental group and are served by SSTE, while another 30\% users are in the control group for the base model.
Different from the evaluation metrics adopted in offline experiments, we introduce three online evaluation metrics that are more concerned in financial recommendation, i.e., total clicks per mille (CLPM), total conversions per mille (COPM) and purchase amount per mille (PAPM).
Specifically, CLPM, COPM and PAPM can be calculated by $\frac{Total \underline{\phantom{x}} Clicks}{Total \underline{\phantom{x}} Impressions} \times 1000$, $\frac{Total \underline{\phantom{x}} Conversions}{Total \underline{\phantom{x}} Impressions} \times 1000$ and $\frac{Purchase \underline{\phantom{x}} Amount}{Total \underline{\phantom{x}} Impressions} \times 1000$, respectively.
The online A/B test results are shown in Figure~\ref{fig:online}.
We can find that our SSTE can bring a steady improvement on the three evaluation metrics.
Overall, our SSTE can achieve an average improvement of 3.75\%, 7.20\% and 12.11\% on CLPM, COPM and PAPM, respectively, in the whole online A/B test.
This further demonstrates the effectiveness of our SSTE.

\section{Conclusions}\label{sec:conclusions}
In this paper, we first show through an online study that blindly removing all biases in an industrial recommendation application may not consistently yield a desired performance improvement.
To achieve a better accuracy-bias tradeoff, we propose a simple yet effective self-sampling training and evaluation (SSTE) framework to preserve the beneficial biases while removing the harmful ones.
Our SSTE contains three new modules, i.e., a self-sampling module constructs debiased subsets for training and validation, a self-training module aims to jointly learn the accuracy-bias tradeoff based on the original training data and debiased subset, and a self-evaluation module aims to capture an optimal model offline based on the original validation data and debiased subsets.
We conduct extensive experiments on a public dataset and a real product dataset, and find that our SSTE can effectively improve the unbiased performance of the recommendation models, and is also of good compatibility.
Finally, our SSTE demonstrates a steady improvement on core evaluation metrics in an online A/B test.

\subsubsection*{Acknowledgements.}
We thank the support of National Natural Science Foundation of China Nos. 61836005, 62272315 and 62172283. 
%
%
%
\bibliographystyle{splncs04}
\bibliography{mybibliography}

\begin{thebibliography}{10}
\providecommand{\url}[1]{\texttt{#1}}
\providecommand{\urlprefix}{URL }
\providecommand{\doi}[1]{https://doi.org/#1}

\bibitem{akiba2019optuna}
Akiba, T., Sano, S., Yanase, T., Ohta, T., Koyama, M.: Optuna: A
  next-generation hyperparameter optimization framework. In: SIGKDD. pp.
  2623--2631 (2019)

\bibitem{bonner2018causal}
Bonner, S., Vasile, F.: Causal embeddings for recommendation. In: RecSys. pp.
  104--112 (2018)

\bibitem{chen2021autodebias}
Chen, J., Dong, H., Qiu, Y., He, X., Xin, X., Chen, L., Lin, G., Yang, K.:
  {AutoDebias}: Learning to debias for recommendation. In: SIGIR. pp. 21--30
  (2021)

\bibitem{he2017neural}
He, X., Liao, L., Zhang, H., Nie, L., Hu, X., Chua, T.S.: Neural collaborative
  filtering. In: TheWebConf. pp. 173--182 (2017)

\bibitem{jadidinejad2021simpson}
Jadidinejad, A.H., Macdonald, C., Ounis, I.: The simpson's paradox in the
  offline evaluation of recommendation systems. ACM TOIS  \textbf{40}(1),
  1--22 (2021)

\bibitem{koren2009matrix}
Koren, Y., Bell, R., Volinsky, C.: Matrix factorization techniques for
  recommender systems. Computer (8),  30--37 (2009)

\bibitem{liang2016causal}
Liang, D., Charlin, L., Blei, D.M.: Causal inference for recommendation. In:
  Workshop on Causation: Foundation to Application co-located with the 32nd
  Conference on Uncertainty in Artificial Intelligence (2016)

\bibitem{lim2015top}
Lim, D., McAuley, J., Lanckriet, G.: {Top-N} recommendation with missing
  implicit feedback. In: RecSys. pp. 309--312 (2015)

\bibitem{liu2020general}
Liu, D., Cheng, P., Dong, Z., He, X., Pan, W., Ming, Z.: A general knowledge
  distillation framework for counterfactual recommendation via uniform data.
  In: SIGIR. pp. 831--840 (2020)

\bibitem{liu2021mitigating}
Liu, D., Cheng, P., Zhu, H., Dong, Z., He, X., Pan, W., Ming, Z.: Mitigating
  confounding bias in recommendation via information bottleneck. In: RecSys.
  pp. 351--360 (2021)

\bibitem{liu2022debiased}
Liu, D., Cheng, P., Zhu, H., Dong, Z., He, X., Pan, W., Ming, Z.: Debiased
  representation learning in recommendation via information bottleneck. ACM
  TORS  (2022)

\bibitem{liu2019spiral}
Liu, D., Lin, C., Zhang, Z., Xiao, Y., Tong, H.: Spiral of silence in
  recommender systems. In: WSDM. pp. 222--230 (2019)

\bibitem{marlin2009collaborative}
Marlin, B.M., Zemel, R.S.: Collaborative prediction and ranking with non-random
  missing data. In: RecSys. pp. 5--12 (2009)

\bibitem{saito2020asymmetric}
Saito, Y.: Asymmetric tri-training for debiasing missing-not-at-random explicit
  feedback. In: SIGIR. pp. 309--318 (2020)

\bibitem{saito2020unbiased}
Saito, Y., Yaginuma, S., Nishino, Y., Sakata, H., Nakata, K.: Unbiased
  recommender learning from missing-not-at-random implicit feedback. In: WSDM.
  pp. 501--509 (2020)

\bibitem{schnabel2016recommendations}
Schnabel, T., Swaminathan, A., Singh, A., Chandak, N., Joachims, T.:
  Recommendations as treatments: Debiasing learning and evaluation. In: ICML.
  pp. 1670--1679 (2016)

\bibitem{swaminathan2015self}
Swaminathan, A., Joachims, T.: The self-normalized estimator for counterfactual
  learning. In: NeurIPS. pp. 3231--3239 (2015)

\bibitem{wang2020information}
Wang, Z., Chen, X., Wen, R., Huang, S.L., Kuruoglu, E.E., Zheng, Y.:
  Information theoretic counterfactual learning from missing-not-at-random
  feedback. In: NeurIPS. pp. 1854--1864 (2020)

\bibitem{wang2022invariant}
Wang, Z., He, Y., Liu, J., Zou, W., Yu, P.S., Cui, P.: Invariant preference
  learning for general debiasing in recommendation. In: SIGKDD. pp. 1969--1978
  (2022)

\bibitem{yang2018unbiased}
Yang, L., Cui, Y., Xuan, Y., Wang, C., Belongie, S., Estrin, D.: Unbiased
  offline recommender evaluation for missing-not-at-random implicit feedback.
  In: RecSys. pp. 279--287 (2018)

\end{thebibliography}
%




\end{document}